# Seismometer Detection of Dust Devil Vortices by Ground Tilt

Ralph D. Lorenz[1*], Sharon Kedar[2], Naomi Murdoch[3], Philippe Lognonné[4], Taichi Kawamura[4],

David Mimoun[3], W. Bruce Banerdt[2]

[1]Johns Hopkins University Applied Physics Laboratory,  11100 Johns Hopkins Road, Laurel, MD 20723, USA.

[2]Jet Propulsion Laboratory, California Institute of Technology, 4800 Oak Grove Drive, Pasadena, CA 91109, USA

[3]Institut Supérieur de l'Aéronautique et de l'Espace (ISAE-SUPAERO), Université de Toulouse, 31055 Toulouse, France

[4]Institut de Physique du Globe de Paris/ University of Paris Diderot, 75205 Paris Cedex 13, France

## Abstract

We report seismic signals on a desert playa caused by convective vortices and dust devils. The long-period (10-100s) signatures, with tilts of ~$10^{-7}$, are correlated with the presence of vortices, detected with nearby sensors as sharp temporary pressure drops (0.2-1 mbar) and solar obscuration by dust. We show that the shape and amplitude of the signals, manifesting primarily as horizontal accelerations, can be modeled approximately with a simple quasi-static point-load model of the negative pressure field associated with the vortices acting on the ground as an elastic half space. We suggest the load imposed by a dust devil of diameter D and core pressure $\Delta P_o$ is ~$(\pi/2)\Delta P_o D^2$, or for a typical terrestrial devil of 5 m diameter and 2 mbar, about the weight of a small car. The tilt depends on the inverse square of distance, and on the elastic properties of the ground, and the large signals we observe are in part due to the relatively soft playa sediment and the shallow installation of the instrument. Ground tilt may be a particularly sensitive means of detecting dust devils. The simple point-load model fails for large dust devils at short ranges, but more elaborate models incorporating the work of Sorrells (1971) may explain some of the more complex features in such cases, taking the vortex winds and ground velocity into account. We discuss some implications for the InSight mission to Mars.

Introduction

A sensitive broadband seismometer, equipped with a wind shield, is presently in development to be emplaced on the surface of Mars by the NASA-led InSight mission, to be launched in 2016. Extensive effort is being devoted to understanding the atmospheric contributions to the seismic signal, since in the absence of microseism-producing oceans, the atmosphere directly dominates the background seismic noise on Mars against which geophysical seismic events must be detected (Lognonné and Mosser, 1993). A planetary surface is not a perfectly rigid structure, and thus it will deform when the loads upon it change. This includes the pressure load exerted by the atmosphere (e.g. Crary and Ewing, 1952; Sorrells, 1971).

A prominent aspect of Mars meteorology is the frequent occurrence of dust devils (e.g. Ryan and Lucich, 1983; Lorenz, 2009), often much larger in diameter (due to the thicker atmospheric boundary layer) than those encountered on Earth. These may play an important role in dust-lifting in the Mars climate system, and cause strong local variations in surface pressure. It seems reasonable to expect that a dust devil may have a seismic signature (on any planet), although there are to our knowledge no reports in the literature of this. While the InSight lander is equipped with a capable meteorology suite which will record pressure, wind, air and ground temperatures in order to decorrelate meteorological contributions from the seismic signal (e.g. Beauduin et al., 1996 ), it may be that seismic instrumentation itself offers a new window on dust devils, and boundary layer convection more generally. Indeed, seismic instrumentation is now being recognized (Pryor et al., 2014) as a useful tool to detect extreme but localized wind gusts on Earth. Moreover, dust devils measured both seismically and in the atmosphere may prove to be a useful tool for calibrating the local elastic properties of the InSight landing site.

We report here on a field campaign wherein a seismometer with a comparable installation to that expected on Mars was deployed on a dry lake bed in California. The emplacement of an instrument on the surface (rather than deeply buried in a borehole) makes it more susceptible to tilt noise (De Angelis and Bodin, 2012); tilt noise due to pressure loads on the surface is also larger on soft ground than on hard bedrock.  One expected outcome of this testing was the observation of seismic data during dust devil encounters, and a partial validation of the noise models used for predicting the seismometer noise due to pressure variations.

Field Measurements

The main goals of the field exercise were to assess 'real-world' effects of a surface deployment of a seismometer on open terrain in a configuration representative of that going to Mars.  Results on such effects as tether noise, thermal effects and lander vibration will be reported elsewhere. While a range of different tests has been performed over 2013-2014, the dust devil investigation here acquired data principally in June 2014, the peak of the dust devil season.

The field measurements were conducted on a ~400m-wide playa about 1km southwest of the Goldstone Deep Space Communications Complex (35°25′36″N 116°53′24″W) outside Barstow, California, within sight of the 70m Deep Space Network antenna (figure 1).  The site was chosen in part due to the existence of a nearby seismic station (CI.GSC, in a vault) and institutional considerations, the access-controlled facility being operated by JPL, which leads the InSight project. In addition the dry lake bed is a good analog to the expected InSight landing site, where a shallow (tens of meters) layer of soft, slow regolith overlays a lava flow.

< Figure 1 – single column fig >

The principal installation (figure 2) comprised data acquisition equipment in a sealed box, powered by a battery and solar panel, logging data at 100Hz from a shallow-buried Trillium compact broadband seismometer.  Additionally, data from an anemometer and a microbarograph were recorded. A small fence was installed to prevent disturbance from wildlife  (in fact wild donkeys are a noted factor in this area).

< Figure 2- single column fig >

The playa surface is fine mud, typically with dessication cracks.  A seismic survey suggested the upper ~5.6m had a seismic velocity of 450m/s, overlying a ~20m thick layer of denser sediment with 750m/s over a faster 1500m/s hard rock (see supplemental information).

For context dust devil information, we deployed eight small self-contained pressure loggers in a cross formation around the seismic station, 30m and 60m distant in each cardinal direction – see figure 1.  This apparatus, used previously for dust devil surveys (see e.g. Lorenz and Jackson, 2014) employed Gulf Coast Data Concepts B1100 loggers (www.gcdataconcepts.com, ~$120), which combine a precision Bosch BMP085 pressure sensor (logged with a resolution of 1 Pa, or 0.01 mb) with a microcontroller that logs the pressure data and housekeeping temperature as ASCII files on a 2GB microSD flash memory card. The whole unit operates as, and its form factor resembles, a large USB memory stick, facilitating data transfer to a PC.   As described in Lorenz (2012), for this application the nominal single AA battery is replaced by a larger battery (in this case two alkaline cells in series), allowing unattended  ~1 month (AA cells) or multi-month (C- or D-cell) operation at sample rates of 2Hz or more. This self-contained power and data acquisition approach is convenient for long-term dispersed measurements and has been used recently to study the horizontal structure of dust devils (Lorenz et al., 2015).  The unit and battery is housed in a plastic case, vented to allow rapid pressure equilibration.

## Identification of Dust Devil Encounters

Since many factors influence seismic signals, our initial dust devil search was performed on the pressure logger data, for which well-established detection criteria exist.  Convective vortices, which may or may not be dust-laden (Lorenz and Jackson, 2014) are detectable as a sharp (typically 10-100s long) dip in local pressure in the time series.  Lorenz and Lanagan (2014) report a survey of such vortex activity at a playa in Nevada : they find a single station encountered about 1 event per day with an observed amplitude of 0.3mbar.   This may be caused by a close encounter of a vortex with a core pressure drop near this value  (Lorenz, 2014), or a more distant encounter with a more intense vortex.  Vortex models (see later) suggest that the pressure drop at the 'wall' is half that at the core, and falls to <10% at about 3 wall radii from the center. The 'wall' radius is that at which the tangential velocity of the vortex has a maximum : in well-defined thinly dust-laden vortices it is in fact visible as a cylinder, as the dust is concentrated here by the balance of the pressure gradient and radial wind drag against the centrifugal force associated with the circular motion.

Since the time-tagging of pressure loggers and seismic data was subject to some uncertainties due to clock drift and operator error, we sought the most conspicuous dust devil events in the pressure data to correlate first. The most prominent example (figure 3), with a distinct close pair of events in the mid-afternoon of June 27[th], 2014, has a peak pressure drop of 0.8 mbar, the other about 0.2-0.8 mbar. The first event is probably the largest identified in the entire campaign. Their characteristic durations (full width half maximum) are about one minute and a little less than a minute, respectively. The ~10 minute spacing is rather typical of dust devils, which tend to form in the edges or corners of a boundary layer convection pattern with a cell size one to a few times the atmospheric boundary layer thickness (~2km) – see Spiga, 2012; Lorenz and Christie, 2015. As this pattern is advected at speeds of a few m/s (in fact the average windspeed recorded by the anemometer at the time was ~7-10 m/s) a 500-1000s interval between devils is often encountered – see e.g. Lorenz and Lanagan (2014).

< Figure 3 – show as full page, portrait>

The seismic disturbance coincides exactly (recorded with the same data acquisition system) with a disturbance noted in the microbarometer at the seismic station. This instrument, like others used in infrasound studies, is not DC-sensitive but its high-pass filtering effectively differentiates signals in the 10-100s band : the dip in pressure due to the vortex (a simple dip is seen directly in the dispersed pressure loggers) therefore manifests as a 'heartbeat' down-up-down signature in the microbarometer (Lorenz and Christie, 2015). Thus we know the seismic event was associated with the vortex passage. The peak accelerations for the E-W and N-S accelerations are ~$6 \times 10^{-7}$ ms$^{-2}$ and $2 \times 10^{-6}$ ms$^{-2}$ respectively for the first event.

The ten-minute spacing between the two events made it easy to correlate the events between the barometers and the seismic station. Both events were seen in all six operating pressure loggers, spaced ~30 and 60m to the North, South and West of the seismometer (one of the two eastern loggers failed altogether; the other had ceased functioning some days previously after several weeks of operation). One operating logger (W1) was equipped with a solar flux monitor, which recorded a brief ~2% dip in sunlight intensity, perfectly coincident with the vortex passage. This confirms that the vortex was dust-laden : dips of about 2% are seen in about 10% of vortex detections (e.g. Lorenz and Jackson, 2015). It should be noted that several of the pressure time series have a double-dip structure : this may suggest that the devil made a slightly cycloidal path with multiple 'close-approaches' or that it had a multiple-core structure. This is, however, a second-order effect.

The encounter was in the evening, with the sun to the west. The drop in sunlight requires that the optical western edge of the devil passed to the southwest of the logger (which itself 30m to the west of the seismometer) in order to cast a shadow on the logger. Since the pressure disturbance is of a similar amplitude (figure 4) for all stations (spanning 100-200m), the diameter of the devil is likely >200m: the duration of ~1 minute, given background winds of ~7m/s, implies a feature of the order of 300m across. Given the wind azimuth just before and after the vortex passage, it is most likely that the vortex moved from south or west. The pressure data alone does not allow a definitive statement on whether the vortex passed to the east or west of the seismic station, and the vortex is large enough that the solar flux data only suggests a slightly higher probability of the center being to the west.

It is possible to constrain the size and miss distance of a dust devil vortex via a wind direction history, as done on Mars by Ryan and Lucich (1983), but unfortunately the sample rate of our windvane was too low to provide useful information.

< Figure 4 – single column fig >

While the first, large event was the easiest to identify, in fact the second, smaller event is easier to interpret. As figure 4 shows, the second signature is large in S1, S2, modest in N1, N2, W1 and negligible in W2. This implies that the feature was comparable in diameter (within a factor of 2 of ~30m) with the miss distances, since the intensity varies between stations. Assuming the devil did not evolve in intensity (dust devil longevity in seconds is typically ~40d$^{0.66}$ (Lorenz, 2014) where d is the diameter – thus a 100m diameter feature should last ~20 minutes) the relative amplitudes imply that it must have moved in a east-northeast direction (heading ~45-75$^o$), passing close to S1 and/or S2. If it moved too far anticlockwise (to the east, 90$^o$) it would have been too small in N1, N2 (and N2 would have been noticeably smaller than N1). If it moved perfectly northwards (heading 0$^o$), it would have been as big in N1, N2 as in S1 and S2. Given the evidence that the station meridian (the N2-S2 line) was crossed between S1 and S2, the stronger signature in W1 than W2 suggests a somewhat north-easterly course, rather than purely eastwards. Similarly, the lack of a sun obscuration signature in W1 suggests the devil was never west-southwest of that station (although of course it could have been a dustless vortex). It is unfortunate that no data from the two east loggers is available as these would have been powerful constraints : nonetheless, the data at hand provide a useful estimate of the intensity (>0.6mb), diameter (between about 20 and 60m) and path of this devil.

It is of interest that a third event is seen in the E-W acceleration history, but is not seen in the pressure loggers or even in the microbarometer record. The fact that the interval between it and the second event is almost exactly the same as between the first two is very consistent with dust devil vortices – similar equispaced triplets were seen in microbarometer records in Australia (Lorenz and Christie, 2015). The fact that none of the pressure loggers sees the third event means it was small in diameter, and must have passed to the East of the seismic station (or it would have showed up in one of the loggers).The fact that the event is seen only in the seismometer record might be taken as an indication that in fact seismic tilt is a particularly sensitive means of detecting these vortices – we will return to this point later.

Quasi-Static Signature of an Isolated Dust Devil

The simplest model of these encounters is the straight-line constant-speed migration of a negative point load on an elastic half-space : in other words, the pressure drop in the dust devil vortex pulls up on the ground at a point. In reality the negative load is a distributed pressure field, but this distinction matters only within a diameter or two of the vortex center. Clearly, the surface will tilt away from the vortex. In the case of a straight-line path directly across the seismometer, the tilt will rise from zero to some maximum value (in practice limited by the separation of the seismometer feet) which then switches sign as the load crosses the instrument and then declines back to zero. In the case of a near-miss, the component of tilt along the direction of motion follows the same functional form, but is muted by the smoother distance history. The component of tilt orthogonal to the direction of motion rises to a maximum value at close approach and declines (but is always of the same sign).

This point model is readily calculated as a function of time using the Boussinesq-Cerruti analytic solutions (derivation from Landau and Lifshitz, 1986). The load to be chosen is determined by integrating the pressure distribution around an analytic vortex model, but clearly should scale with $\Delta P_o D^2$, i.e. the area and core pressure drop, where D is the wall diameter. In fact, the two vortex models discussed by Lorenz (2014), one by Vatistas (1992) with $\Delta P(x)=\Delta P_o[1-(2/\pi)$ arctan $(r^2)]$ where $r=2x/D$ and $\Delta P(x)$ is the pressure drop observed at distance x from center, and a Lorentzian form $\Delta P(x)=\Delta P_o(1/[r^2+1])$ used by Ellehoj et al. (2010) to model Martian dust devils, fail in this application. While they adequately describe near-field pressure data, they do not fall off fast enough with distance to converge to a finite load : for a finite integral, the pressure must fall off faster than $1/r^2$. Using a steeper function, such as $\Delta P(x)=\Delta P_o(1/[r^3+1])$ and summing the incremental load on a ring dx wide, $2\pi x\Delta P(x)dx$, to infinity, a total load of $1.95(\pi/4)\Delta P_o D^2$ is obtained, i.e. just over double that of a disc equal in diameter to the wall of the dust devil, uniformly loaded at the core pressure drop. Field data (Lorenz et al., 2015) and laboratory data (Vastitas et al., 1991) do not strongly discriminate these candidate functions, but the $r^2$ dependence is probably closer to the truth : truncating the arctan($r^2$) function at r=5 gives

~3$(\pi/4)\Delta P_o D^2$. If we instead adopted an r exponent of 2.5 and integrated to infinity the prefactor would be ~1.1 – the final result is therefore not very sensitive to the exact function used, and for convenience, then, we adopt the succinct intermediate expression L=$(\pi/2)\Delta P_o D^2$ as our nominal result. For a typical small terrestrial dust devil, D=5m, $\Delta P_o$ = 2mbar, we then find a load L=7900 N, or roughly the (negative) weight of a small car.

The predicted pressure variation and the seismic tilt of such a typical terrestrial dust devil is shown in Fig. 8 for several miss distances.   This simulation evaluates tilt using an assumed foot separation of the seismometer of 20.3 cm (the feet are spaced on a circle of 23.5cm diameter), and a Young's modulus E of the playa ground of 337 MPa, with a Poisson's ratio of 0.25.  E is chosen to be consistent with a measured seismic P-Wave velocity of 450m/s, assuming a bulk density of 2000 kg/m$^3$ (see Supplemental Information).

When the closest approach distance is equal to 10 m (i.e. r=4), a 5 m diameter devil will cause a 0.03mb pressure drop and a maximum tilt of ~$7 \times 10^{-7}$ m/s$^2$   (i.e. 70ng acceleration).   As described qualitatively above, the component about the line to closest approach has a 'heartbeat' signature, changing sign as the devil passes by, while the orthogonal axis sees a rise then a fall. Clearly, for an arbitrarily-oriented seismometer, the observed signatures would be linear combinations of these histories weighted by the sine and cosine of the azimuth of close approach in the seismometer reference frame (and this may account for the shape of the first encounter in figure 3).

< Figure 5 – full page portrait >

It is seen in figure 5 that the pressure history is more strongly dependent on miss distance than is the tilt history.   In fact, since we assume the dust devil load is applied only vertically, and we consider only the vertical ground displacements under the seismometer, the maximum tilt observed varies with KL/x$^2$, L being the load, x being the closest approach distance and K being a constant that describes the ground: $(1+\nu)(1-\nu)/\pi E$, where $\nu$ is the Poisson's ratio and E is the Young's modulus. The resulting maximum tilt is proportional to $K\Delta P_o D^2/x^2$. Thus, while the tilt signature is inversely proportional to $x^2$, the pressure perturbation falls off as greater than $x^2$ and thus seismometers may be effective at detecting vortices at longer ranges than are pressure sensors We note, however, that better pressure data are needed in the far field of real devils to know what the correct dependence on observed pressure with distance should be.

Since the first event observed had a duration of ~1 minute, and caused very similar pressure perturbations on loggers spaced ~90m apart, it seems it must have been rather large, >>100m. The core pressure drop may well have been ~2 mb – certainly more than 1mb. Thus its total loading may have exceeded $3x10^6 N$ (i.e. 300 tons), or 400 times the 'typical' devil above. Assuming a miss distance of 50-200m (the $x^2$ factor reducing tilt by 25-400 relative to the 10m encounter), the peak tilt for the observed events should be of the order of $6.8x10^{-7}$ to $1.1x10^{-5}$, $m/s^2$ as observed. An example fit, found by least-squares error on the tilt signatures, with an initial guess guided by the interpretation of the pressure logger records, is shown in figure 6. While the overall shape and magnitudes are indicated reasonably by the model, taking the pressure logger information into account in choosing the size and trajectory, this first event (see figures 3 and 6) has a somewhat complicated structure, evident in the pressure records that suggest it either had a somewhat cycloidal migration path or had multiple vortex cores. The large size of the vortex compared with the miss distance also challenges the basis of the simple point-source model, so we do not attempt more elaborate fitting procedures.

< Figure 6  - show as full page landscape >

Considering the second event, we choose a couple of example encounter geometries, guided by the pressure logger information described in section 3.  There is some degeneracy between distance and time (in that a large structure advected quickly will give a similar curve shape to a small one advected more slowly – see figure 5), but of course a large structure has a higher total load.

Comparing the two example fits in figure 6, we see that the change in heading from East-south-east (110$^o$) to East-north-east (~80$^o$) produces a significant change in the shape of the E/W tilt signature. A better fit is obtained in the former case, but this is circumstantially less consistent with the pressure logger data. Given the uncertainties in the far-field fall-off of pressure with distance, and the possibility of intensity evolution and/or curved migration paths, we do not attempt a global unified fit of both pressure logger and seismometer signals, but do conclude that the tilt signature appears to be well-captured by the simple model in this instance. We note that none of these fits are unique, and also that accurate time-tagging of the pressure loggers would greatly assist reconstruction efforts.

We have considered only the quasi-static response to a point load being advected at constant speed. The vortex winds themselves will apply loads to the ground, and close encounters with dust devils (which lead to sudden swings in wind direction) may therefore see tilt variations (and, indeed, velocity signals, since the tilt changes rapidly – see also Sorrells, 1971; Sorrells et al., 1971) as a result of the elastic deformation of the ground. Further, the ground is not a perfectly elastic solid. Sorrells' theory pertains to a straight-line front at which there is a step change of pressure – this is arguably a better description of a large vortex in the near-field than is the point model, and his theory also includes the effect of wind. We note also that the ground deformation leads to a vertical movement of the seismometer

## Implications for Mars

The Mars Pathfinder and Phoenix missions both carried pressure sensors sampled at an adequate rate to identify pressure drops as likely dust devil encounters. Around 1-4 vortex encounters were detected per day (Lorenz et al., 2012) with thresholds of 30 μbar for Phoenix (Ellehoj et al., 2010), which had the larger number of detections, and 50 μbar for Pathfinder (Murphy and Nelli, 2002). Both datasets show a similar -2 power-law frequency dependence on peak pressure drop.

Because the atmospheric pressure (and thus density) on Mars is a couple of orders of magnitude smaller than on Earth, a dust devil has a smaller absolute pressure dip (although encounters of a given relative pressure dip may be slightly more frequent – see Lorenz and Lanagan, 2014). Thus a dust devil of a given diameter on Mars will exert a weaker load on the surface; however, this is partly compensated by the larger typical diameter of Martian dust devils.

If we adopt a ~15m diameter and a core drop of 100 μbar as a large but not exceptional Mars lander encounter, the total load is similar to the typical terrestrial value of ~7000N, and a close encounter 30m away would lead to a sensed pressure drop function) of >10 μbar – typical of events seen by the Phoenix lander almost daily. If the Martian regolith beneath InSight has the same response as our playa mud, such a 30m encounter will yield a tilt of ~5 nanoradians, easily detectable by the InSight SEIS instrument, which has a specified noise level on its horizontal axes of $10^{-9}$ m/s$^2$/($\sqrt{}$Hz) in the 0.1 to 1 Hz bandwidth (Lognonne et al., 2012). It seems probable that many dust devil signatures will be encountered.

Conclusions

We have reported for the first time the detection of dust devil encounters with a seismometer in a field experiment, documented in part by an array of pressure loggers. The characteristic tilt histories observed are consistent with the passage of a negative load associated with dust devil vortices, and the amplitudes appear consistent with reasonable vortex parameters, and a simple point-load model appears to adequately describe small/distant encounters. Larger vortices demand a more sophisticated approach, and we suggest the theory of Sorrells (1971) may be promising in its ability to recover the structure of the tilt we observe in the case of a large vortex.

We will construct more elaborate models in future work. First, the point model is inaccurate when the encounter distance is small compared with the dust devil diameter (as in the first encounter we report here), since a significant part of the pressure field acts on the opposite side of the seismometer from the devil center. For a homogenous elastic half-space model, one could decompose the pressure field into an array of incremental point loads and sum the tilt contributions (qv Dunkin and Corbin, 1970). A full finite-element study could be employed (e.g. Kroner et al., 2005). Further, the response of the ground to the time-varying pressure load following Sorrells (1971) but applying a realistic model of vortex winds (like a Rankine vortex, or the more physical Vastitas et al 1990 formulation) will be needed, rather than the plane wind in Sorrells' model.

A key assumption in the simple models is that the dust devil migrates with a constant velocity, whereas field observations and the dramatic sweeping curls of dust devil tracks on Mars attest to meandering and sometimes cycloidal migration paths: these can lead to pressure histories that have complex shapes and multiple dips (e.g. Lorenz, 2011).

A seismometer appears to be capable of tracking close encounters with dust devils, recovering an estimate of the azimuth history and constraining the integral of the pressure field (relating to diameter and core pressure drop). In combination with wind and single point pressure measurements – if the wind data are acquired at a high enough cadence – the dust devil parameters and miss distance may be reconstructed or at least constrained.

While this paper has examined in detail only the low-frequency tilt signature of a dust devil, there are high-frequency components to both pressure and seismic signals (see figure 3). Tatom et al. (1995) suggest that seismic observations might give early warning of tornado encounters, and cite a number of anecdotal descriptions of ground vibration noticed by observers. While these may be due in part to side loads on trees and buildings (not present on terrestrial playa, or on Mars), there may be azimuthal variations (such as multiple cores), which could produce some high frequency seismic or infrasound signals.

The approach described here considers dust devils as discrete entities (certainly the impression one gets visually in the field) but in fact they are merely the most intense of a whole spectrum of turbulent pressure loads associated with the convecting boundary layer : upwelling sheets of air will apply pseudo-line loads which while having smaller local pressure drops than dust devil vortices, may have much larger areas : modeling of boundary layer convection on Mars is therefore of interest to estimate the background noise. As noted by Sorrels et al. (1971) and Douze and Sorrells (1975), much of the seismic noise at a station is correlated with the pressure history, which can be used to estimate and therefore remove that noise (Beauduin et al., 1996; Lognonne and Mosser, 1993). The strong pressure gradients in dust devils make it likely that noise in close encounters cannot be completely decorrelated in this way, however, but dust devils themselves are interesting objects of study, and may act (at least in aggregate) as a set of calibration loads with which to infer the elastic properties of the regolith at the InSight landing site.

## Data and Resources

Data presented in this paper may ultimately be released on the NASA Planetary Data System (PDS) as part of the calibration dataset for the InSight mission. Pending such release, the authors may be able to make the data available upon request.

## Acknowledgements

RL acknowledges the support of the NASA Mars Fundamental Research and Mars Data Analysis programs, via grants NNX12AJ47G and NNX12AI04G. The results reported here benefited from the contributions of many individuals at JPL, IPG, Goldstone, UCLA and Caltech: we particularly acknowledge Paul Davis and Rob Clayton for loan of equipment and determination of the seismic properties of the playa. We thank Jim Murphy and an anonymous referee for constructive comments.

Mailing Addresses

Johns Hopkins University Applied Physics Laboratory,  11100 Johns Hopkins Road, Laurel, MD 20723, USA  (RL)

Jet Propulsion Laboratory, California Institute of Technology, 4800 Oak Grove Drive, Pasadena, CA 91109, USA    (SK, WBB)

Institut Supérieur de l'Aéronautique et de l'Espace (ISAE-SUPAERO), Université de Toulouse, 31055 Toulouse, France  (NM, DM)

Institut de Physique du Globe de Paris/ University of Paris Diderot, 75205 Paris Cedex 13, France (PL, TK)

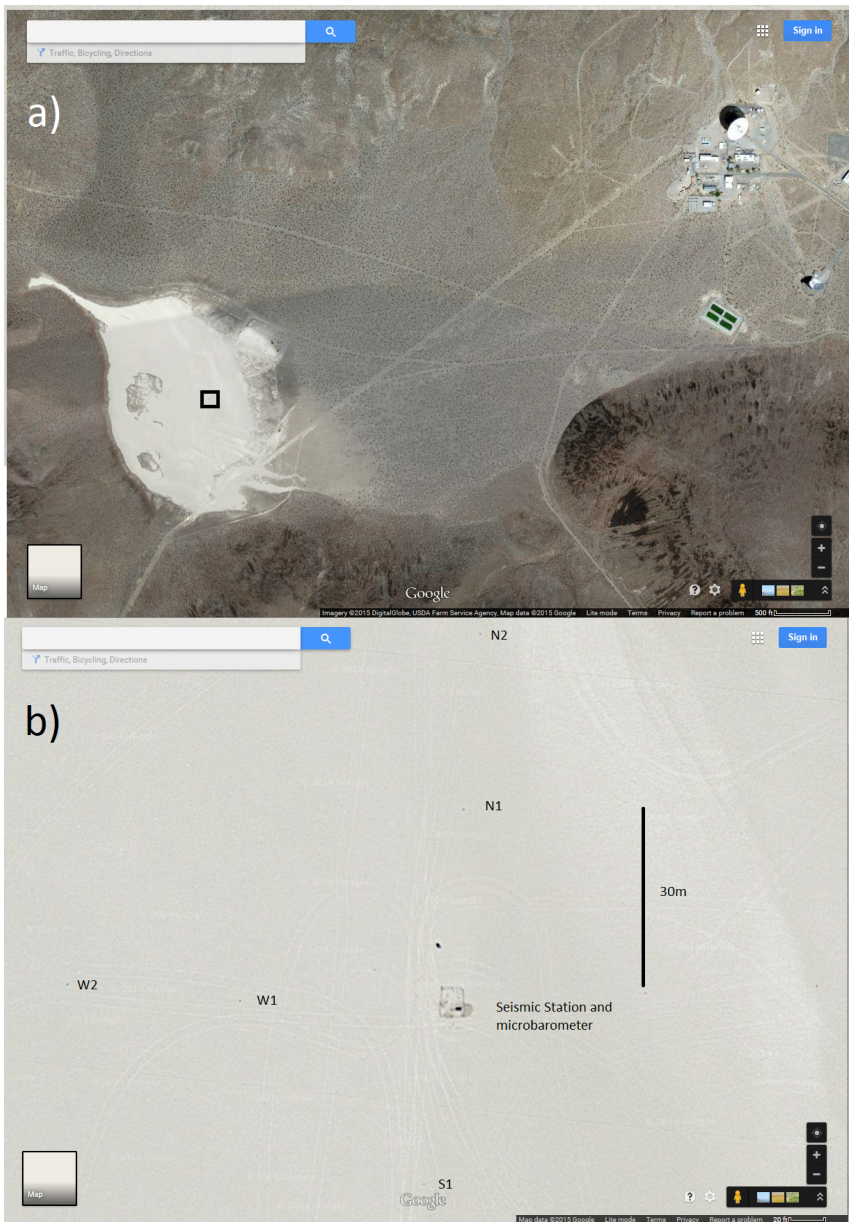

Figure 1. (a) Satellite image showing the Goldstone deep space communications complex (note the prominent white 70m dish and its shadow) at upper right and the playa at lower left. The black box denotes the region zoomed in (b). (b) close-up of the field installation – the seismic station and its small corral is seen at the center, and the small dark dots are pressure loggers arranged in a north/south east/west cross. Logger S2 is below the bottom of the image. Some vehicle tracks are faintly visible

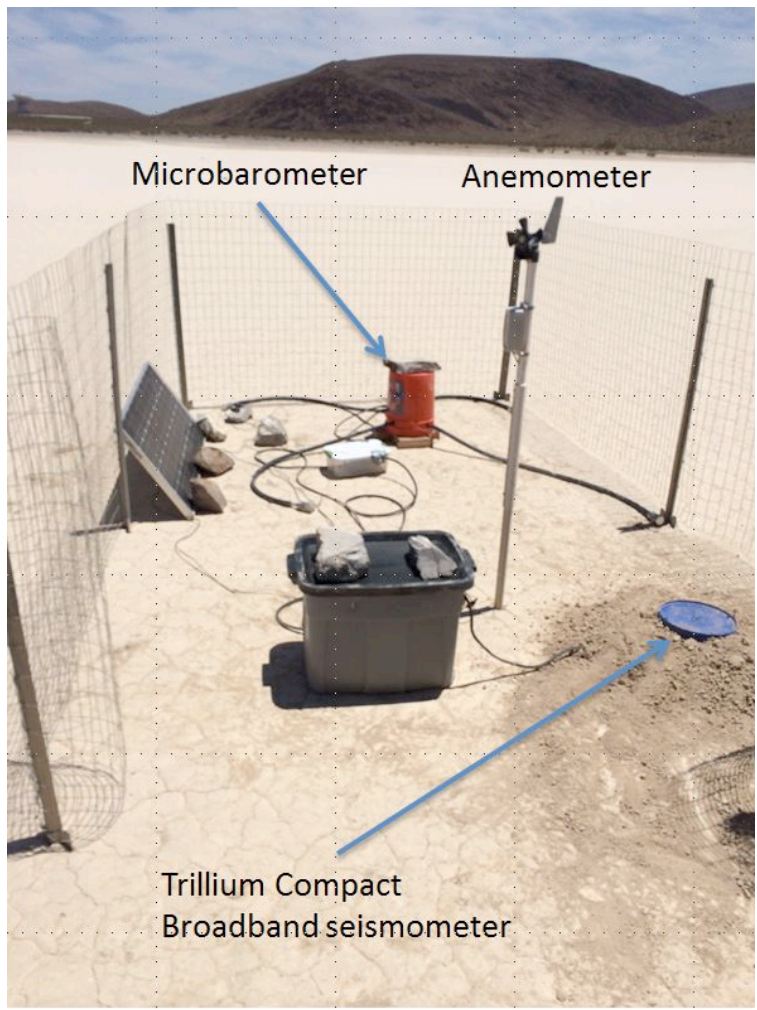

Figure 2. Field installation on the playa including a Nanometrics Trillium Compact seismometer, an Anemometer, and a MB 2005 Microbarometer.

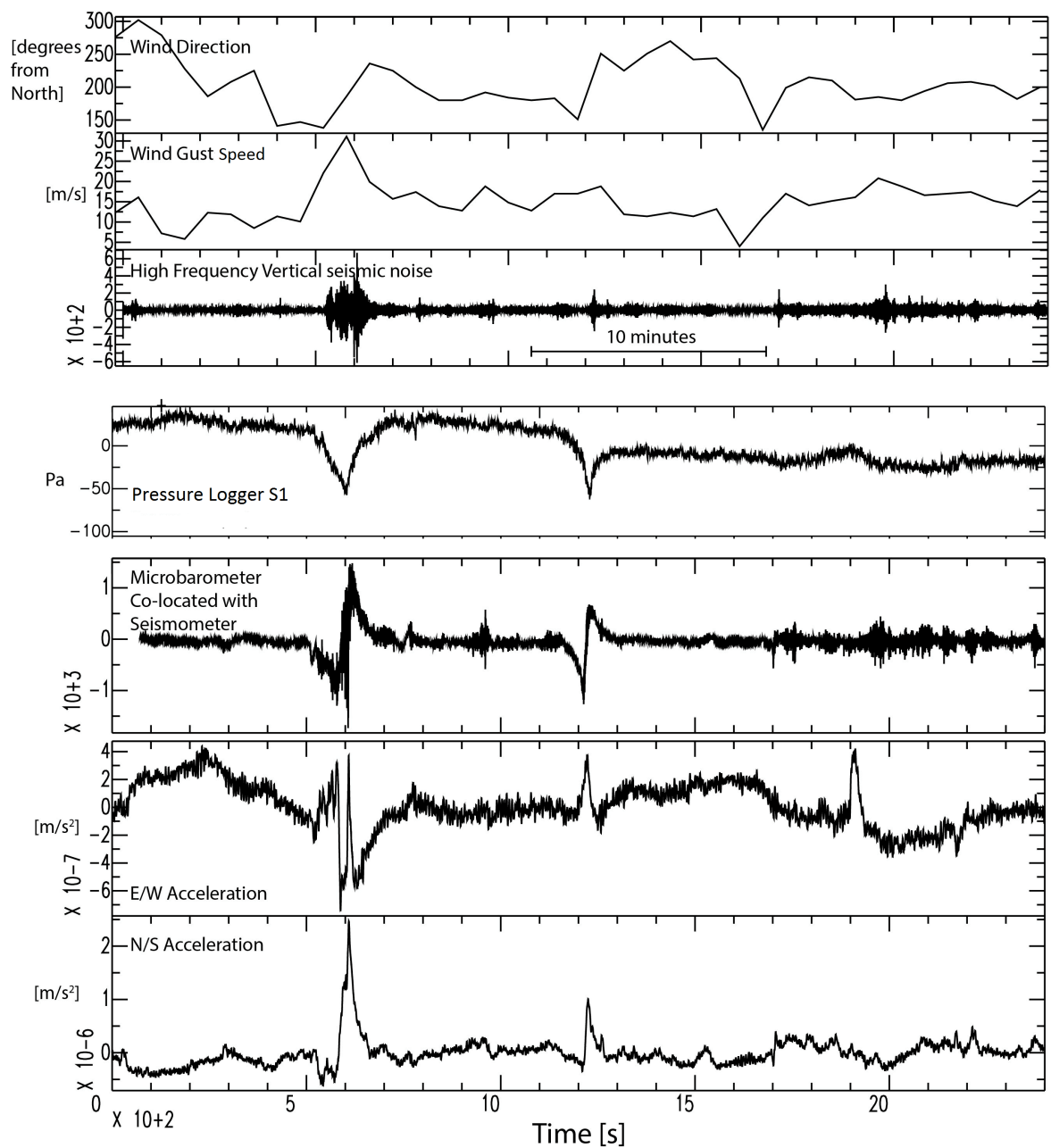

Figure 3. A distinctive pair of events seen at 23:13 and 23:23 UT (i.e. mid-afternoon) on 27th June 2014. The average wind direction over 1 minute, and the peak 10-s gust speed within that minute, is shown in the upper panels. Pressure loggers (e.g. S1, shown here) could not be formally synchronized with the other data (nor are they co-located) but this example been shifted arbitrarily to align with the seismometer events : the correspondence of the two events in all the datasets  (the microbarometer record which was both co-located and perfectly synchronized with the seismic data)  is a differentiated version of the pressure logger record) demonstrates the clear association of a tilt signature with the passage of dust devils.. A third event is apparent in the E-W tilt signal only, ten minutes after the second.

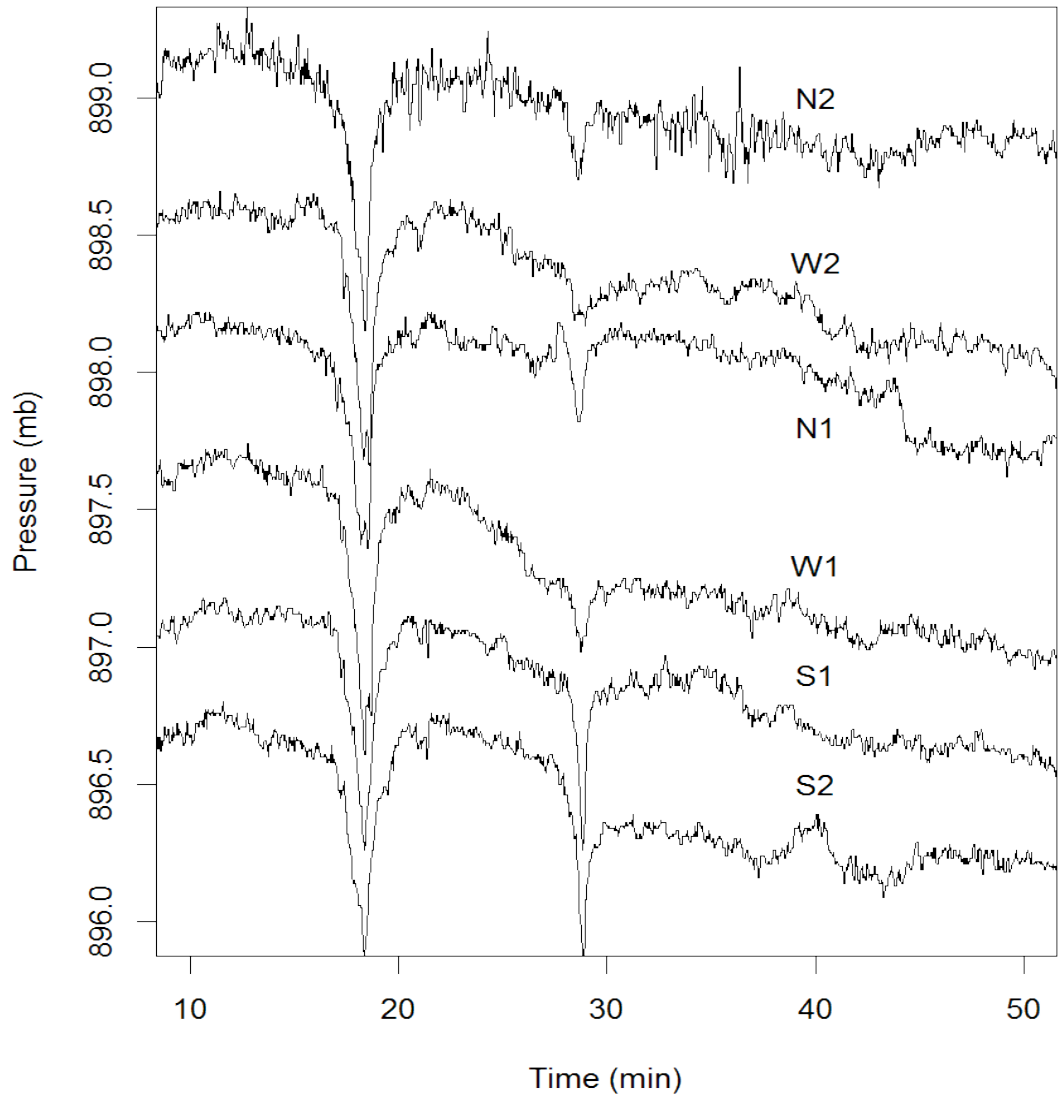

Figure 4. Pressure logger records of the dual (triple) event. The curves are offset vertically for clarity. Since it was not possible to synchronize the records to better than a couple of minutes no trajectory information can be recovered from timing, so the times have been adjusted to match exactly. Note that the first event is of a similar magnitude in all loggers (implying an extent wider than the separation of the loggers) whereas the second, smaller event is strong in S1,S2, weaker in N2,N1 and W1, and very weak in W2. The implications for diameter and trajectory are discussed in the text.

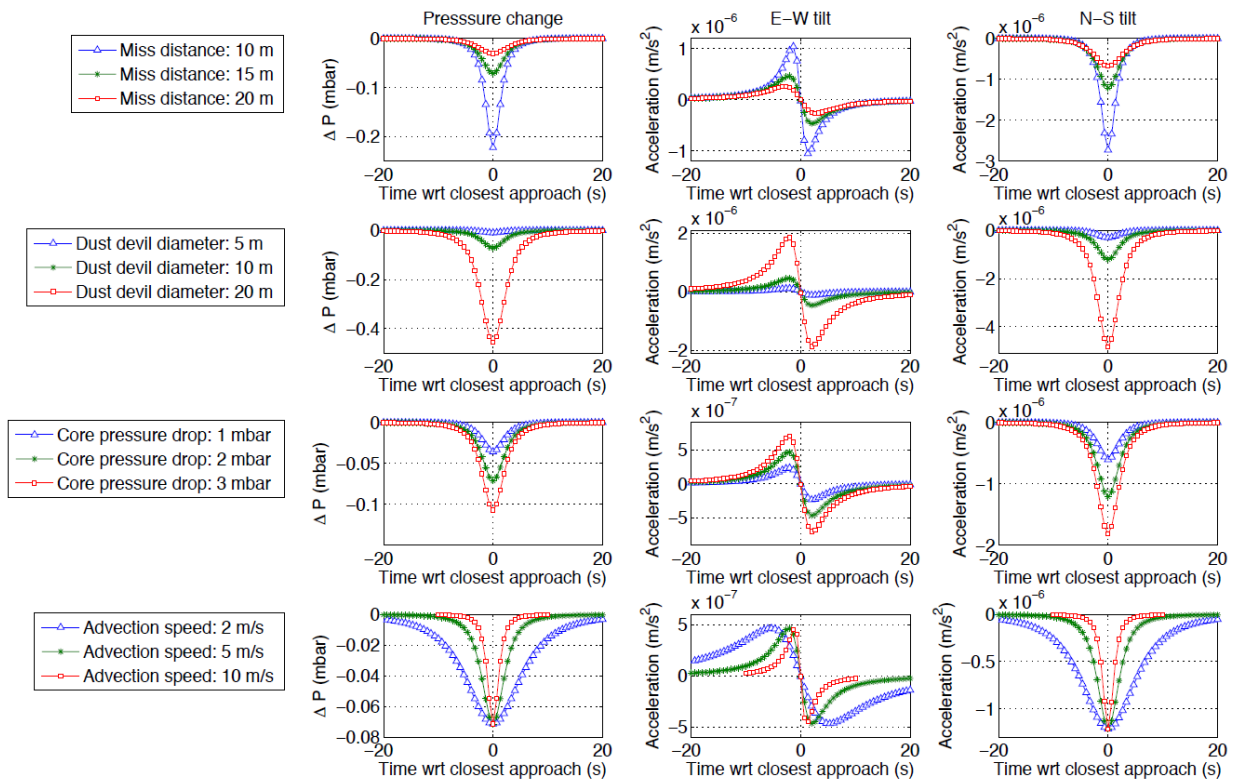

Figure 5. The theoretical pressure variation observed at the seismometer and the horizontal acceleration measured by the seismometer (in the E-W and the N-S direction) due to passage of a dust devil to the north of the station (thus negative N-S tilt means ground tilts towards the south), moving West to East. The effect of varying the different parameters is seen : when not otherwise specified, the advection speed is 5 m/s, the core pressure drop is 200 Pa (2 mbar), radius: 5 m, and closest approach/miss distance: 15 m. A larger diameter and lower speed broadens the signatures ; larger diameter and core pressure drop magnify the signatures. An arbitrary migration direction would mix the E-W and N-S signature components.

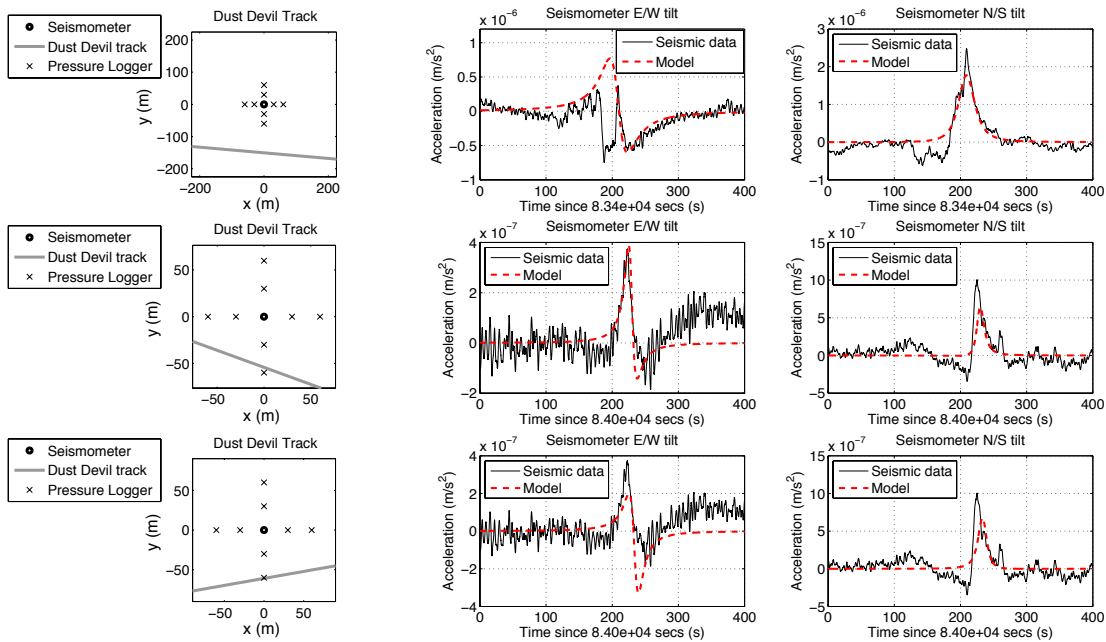

Figure 6. Model comparison with the seismic events. (Upper figures): Modeled seismic tilt compared with the field data for the first event. Advection speed: 8 m/s at heading 95°, Core Pressure drop: 150 Pa, Radius: 70 m, Closest approach/miss distance: 150 m. The overall characteristics of the seismic signature are reasonably produced by the model, but the structure of the E-W tilt history is not completely captured. (Middle figures): Modeled seismic tilt compared with the field data for the second event. Advection speed: 6 m/s at heading 110°. Core Pressure drop: 80 Pa, Radius: 20 m. Closest approach/miss distance: 51 m. (Lower figures): An alternative fit to the second event, showing the effect of a different heading, which significantly changes the shape of the E-W tilt signature. Advection speed: 6 m/s at heading 80°. Core Pressure drop: 50 Pa, Radius: 30 m, Closest approach/miss distance: 60 m.